\documentclass[aps,twocolumn,prl,showpacs,superscriptaddress]{revtex4}
\usepackage{graphicx}
\usepackage{epsfig}
\usepackage{amsmath}
\DeclareMathOperator{\tRe}{Re}
\begin{document}
\title{ {\bf A simple model for self organization of bipartite networks}}
\author{Kim Sneppen}
\affiliation{NORDITA, Blegdamsvej 17, 2100 Copenhagen {\O}, Denmark}
\author{Martin Rosvall}
\affiliation{Department of Physics, Ume{\aa} University, 90187 Ume{\aa}, Sweden}
\affiliation{NORDITA, Blegdamsvej 17, 2100 Copenhagen {\O}, Denmark}
\author{Ala Trusina}
\affiliation{Department of Physics, Ume{\aa} University, 90187 Ume{\aa}, Sweden}
\affiliation{NORDITA, Blegdamsvej 17, 2100 Copenhagen {\O}, Denmark}
\email{trusina@tp.umu.se}
\author{Petter Minnhagen}
\affiliation{NORDITA, Blegdamsvej 17, 2100 Copenhagen {\O},
Denmark} \affiliation{Department of Physics, Ume{\aa} University,
90187 Ume{\aa}, Sweden}

\date{\today}

\begin{abstract}
We suggest a minimalistic model for directed networks
and suggest an application to injection and merging of 
magnetic field lines.
We obtain a network of connected donor and acceptor vertices
with degree distribution $1/s^2$,
and with dynamical reconnection events of size $\Delta s$ 
occurring with frequency that scale as $1/\Delta s^3$.
This suggest that the model is in the same universality class as the
model for self organization in the solar atmosphere
suggested by Hughes et al. \onlinecite{hughes}.
\end{abstract}


\maketitle

In a number of physical systems one observes 
emergence of large-scale structures, caused
by growth of small-scale fluctuations.
For example, 1) the energy flows from small to large scales in 
2-d turbulence, 2) the matter distribution in the universe is
highly inhomogeneous in spite of a presumably uniform energy distribution
at its origin, and 3) the magnetic field lines reconnection and sunspot
activity is able to generate solar flare activity with burst sizes that
by far exceed excitations associated to the individual
convection cell on the solar surface.
In fact, often the emerging large-scale structures exhibit 
scale-free features over substantial range of scales,
as e.g. the sun spots \cite{close,paczuski}
and solar flare activities \cite{ashwanden,parker,hughes}. 

Recently it has been realized that many complex networks
exhibit scale-free
topologies \cite{Faloustos,albert,Broder}, including
in particular the topology of sun spots connected by magnetic field lines
\cite{close,paczuski}.
In general, the first theoretical framework for
emergence of power law distributions was the Simon model \cite{simon},
featuring  a ``rich get richer'' process, that recently has been
developed into \emph{preferential attachment} to explain
scale-free networks \cite{albert}.
An alternative approach to generate large-scale
features from small-scale excitations is provided by the
self organized critical (SOC) models 
\cite{BTW87,forest-fire,BS93} which in their traditional 
versions propose a scenario for 
the fractal pattern of activity that is observed in systems with 
extreme separation of timescales. 
Hughes et al. \cite{hughes} has proposed 
a SOC like mechanism for cascades of reconnection
of magnetic field lines in the solar atmosphere, using a plausible
number of processes associated to diffusion of sun spots and 
reconnection of crossing field lines.
In this paper we suggest a simpler model, 
assuming only two processes, merging and creation,
in an on going dynamics of vertices connected in a network.
\\\\

\begin{figure}
\centerline{\epsfig{file=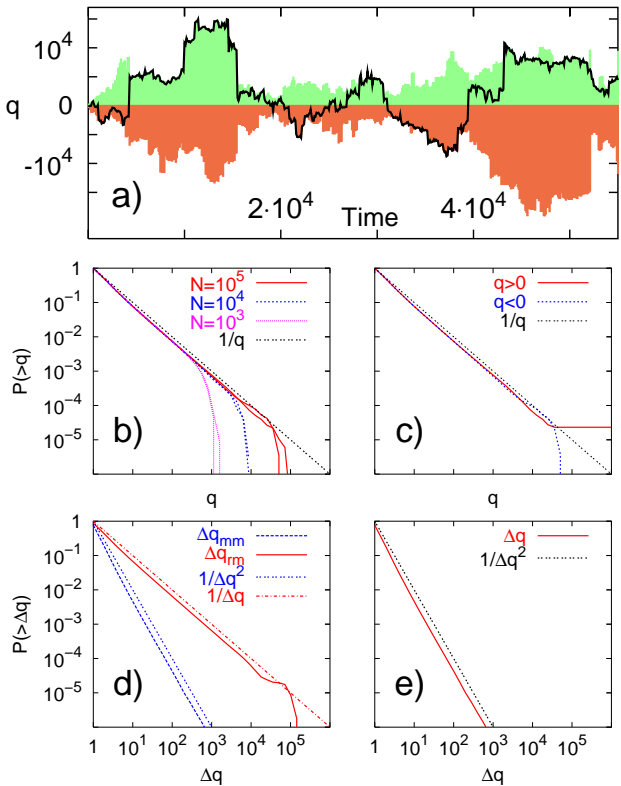, height=0.45\textheight,angle=0}}
\caption{Main features of the basic model.
In panel {\bf a)} we show the development of 
the elements with largest positive
 $q_{max}$  and largest negative $q_{min}$  together with a 
particular trajectory, where we always follow the ``winner" in the 
merging process (solid line). The simulation is done for 
a system of size $N=10^4$ elements,
and the time-count is in updates per element.
b-e) Cumulative plots of steady state properties. 
{\bf b)} The size distributions of the positive and negative 
$q$ for 3 different system sizes.
{\bf c)} The size distribution when we start a system of size
$N=10^5$ with initial condition $q_i=10$, $\forall i$.
One observes that all excess $q$ is moved to a single element.
{\bf d)} Two variants of a histogram of size changes.
$\Delta q_{mm}$ is defined by following the winner,
 meaning that we plot the difference in
size between the largest of $q_i,q_j$ before merging and the 
merged unit after the merging.  $\Delta q_{rm}$ corresponds 
to the change from any of the two $q_i,q_j$ to the merged unit.
{\bf e)} The size of changes defined as the losses of absolute $q$
in merging events where $q_i$ and $q_j $ are of different signs.
}
\label{pownet}
\end{figure}

We first review the basic process of merging-and-creation
( originally proposed by \cite{takayasu,takayasu2})
in a formulation that is closest to the network interpretation
which we will discuss later.
The model describes the evolution of a system of many elements 
$i=1,2,....,N$ that each is characterized by a
scalar $q_i$ that may be either positive or negative. 
One may think of the scalar as a helicity or as 
a quantification to which extent an element/vertex 
is a donor or an acceptor. The model
describes a situation in which the elements in the system
redistribute their respective charges $q_i$ according to
\begin{eqnarray}
merging:\;\;\; q_i &\rightarrow & q_i + q_j
\nonumber\\
q_j & \rightarrow & 0
\label{merging}\\
creation:\;\;\; q_k & \rightarrow & q_k+1
\nonumber\\
q_l & \rightarrow & q_l-1
\label{creation}
\end{eqnarray}
With $(k,l)$ selected independently from $(i,j)$ 
these two processes define one of the many possible realizations of the model.
Other realizations include different combinations of correlations between 
$(k,l)$ and $(i,j)$.
For example, one may select $k=j$ and $l=i$. 
For any choice the obtained scaling is as reported 
in Fig.\ \ref{pownet}.

The main features obtained numerically are presented 
in Fig.\ \ref{pownet}.
Fig.\ \ref{pownet}(a) 
illustrates the steady state after a transient time $\sim N$
updates per element, starting from an initial ``vacuum" with   
$q_i = 0, \forall i=1,2,...N$.  The figure shows the extreme 
range of $q$-values at any time.
The subsequent dynamics of the extremes is also reflected in 
the trajectory of a winner-element which, when merged, 
is re-identified as the merged element.
One observes that this winning element exhibits an intermittent dynamics
with size-changes $\Delta q$ of all magnitudes.
The distribution of these changes as well as a wide set of other properties
is in fact scale invariant.
The cumulative distribution of $q$-values, Fig.\ \ref{pownet}(b), 
is a scale-free distribution, 
\begin{equation}
P(>\!\!q) =\int_q^{\infty} P(q') \mathrm{d}q' \propto q^{1 - \gamma},
\end{equation} 
with $\gamma=2$.
With asymmetric initial condition, say 
$q_i = 10, \forall i=1,2,...N$, as illustrated in Fig.\ \ref{pownet}(c), 
the system self-organizes by concentrating all of the 
initial asymmetry to one of the elements. 
All other elements  are distributed in exactly the same way as 
with the ``vacuum" initial condition 
(compare Figs.\ \ref{pownet}(b) and \ref{pownet}(c)). 

Fig.\ \ref{pownet}(d) shows the distribution of changes 
in $\Delta q$ under steady state 
conditions. There are two possible ways to characterize these changes.
One may quantify them by considering the difference between the 
merged element ($q_i+q_j$) and any of the two 
$q_i$ or $q_j$ merging elements.
In that case one observes a cumulative distribution
for changes $P(>\!\!\Delta q) =1/(\Delta q)$, 
(compare full drawn line and dashed line in Fig. \ref{pownet}(d)).
This distribution closely resembles the overall
distribution of $q$ values.
Alternatively one may 
quantify the dynamics by following the winner at each merging, 
thus defining the $\Delta q$ as the difference between the largest
$q$ before and the largest $q$ after the merging.
In that case one expects the probability of change of size $\Delta q$ 
\begin{eqnarray}
P_{change}(\Delta q) = & & P(q_i = \Delta q) \cdot P(q_j>\Delta q) \nonumber\\
& + & P(q_j = \Delta q) \cdot P(q_i>\Delta q) \nonumber \\
\propto & & \frac{1}{\Delta q^{\gamma}} \frac{1}{\Delta q^{\gamma-1}} \; = \; 
\frac{1}{\Delta q^{2\gamma-1}},
\label{eq2}
\end{eqnarray}
which with $\gamma=2$ from Fig.\ \ref{pownet}(b) predicts exponent $3$ 
verified by simulations, see Fig.\ \ref{pownet}(d). 
Finally  Fig.\ \ref{pownet}(e) shows the size-distribution of annihilation
events, defined as events where two elements of different signs merge.
The distributions of these annihilations are governed by
the same considerations as in Eq.\ \ref{eq2}, and accordingly
scales with exponent $\tau=3$.

Now we explore the reason for the $\gamma=2$ scaling behavior.
We consider the version of the model where one excites the system 
by randomly picking a zero element and assigning it a $+/-$ value: 
\begin{eqnarray}
q_i & \rightarrow & q_i+q_j\\
q_j & \rightarrow & r
\end{eqnarray}  
were $r$ is a random number picked from a symmetric narrow 
distribution $F(q)$.
This update is one of many possible versions
that all produce the same scaling results as shown in Fig. 1,
we here consider it because it is the simples to treat analytically.
The differential equation, describing the evolution of the model reads
\cite{takayasu,takayasu2} 
\begin{eqnarray}
\frac{dP(q)}{dt} = & & 
\iint_{\!\!\!-\infty}^{\infty}\mathrm{d}q_2 \mathrm{d}q_1 \delta(q-q_1-q_2) P(q_1)P(q_2) 
\nonumber \\
 & - & 2P(q) + F(q),
\label{eq3}
\end{eqnarray} 
which have been shown to give a steady state distribution
with the asymptotic behaviour $P(q)\propto 1/q^2$
\cite{takayasu}.
For pedagogical reasons we here present an alternative solution,
that also opens for some insight into the amazing robustness 
of this model.
In terms of the Fourier transform 
$p(\omega) =\int \mathrm{d}q e^{-i q\omega} P(q)$  the steady state 
equation is
\begin{equation}
p(\omega)=1-\sqrt{1-f(\omega)}.
\end{equation}
The important
property is that $p(\omega)-1\propto -|\omega|$ for small $\omega$.
A positive creation probability $F(q)$ 
with a finite second moment ensures this which leads to
$P(q)\propto q^{-2}$ for large $q$.
Thus the exponent $2$ will be a common property for a large
class of variations of the basic merging and creation mechanism.
As an example $F(q)=\exp(-|q|)/2$ gives
\begin{eqnarray}
P(q)=\frac{1}{\pi}-\frac{1}{\pi}\tRe [S_{11}(iq)] = 
\frac{1}{\pi q^2}-\frac{3}{\pi q^4} + \cdots,
\end{eqnarray} 
where $S_{11}(z)$ is a Lommel function and $P(k=0)=1/\pi$.
Also the localization of positive excess $\langle q\rangle N$ 
can be understood, since
a symmetric $F(q)$ implies an even continuum solution and thus that
all excess will occupy a zero $q$ measure around $q_0= \langle q\rangle N$. 
For a discrete simulation this
means a single $q$ as illustrated in Fig. \ref{pownet}c.

\begin{figure}
\centerline{\epsfig{file=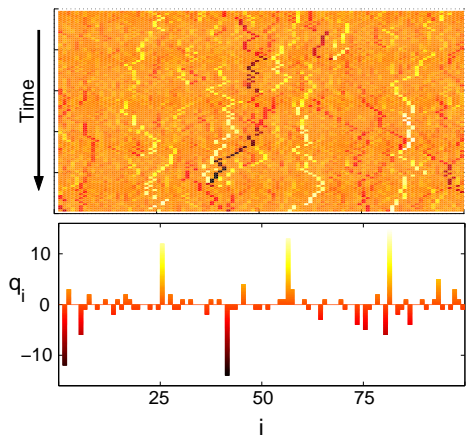, height=0.25\textheight,angle=0}}
\caption{Merging and creation in 1-dimension:
At each time-step one selects a coordinate $i$ between 1 and N=100.
If $q_i=0$, one create a $+/-$
pair at position $i$ and one of its neighbors.
If $q_i\neq0$ then one moves $q_i$ one step to either left or right
and adds it to the $q$ already present at that position.
In upper panel one sees the time evolution over in total 100 updates per
site, whereas the lower panel is a snapshot of the final configuration.
One may notice that merging of light (positive) and dark 
(negative) sites leads to annihilation of both.
The steady state distribution of $q$ is scale invariant with 
the same exponent 2 as in the basic (infinite dimensional) model.
}
\label{1d}
\end{figure}

Finally, for application in real physical situations, 
it is also of interest to explore the behaviour of the 
merging-and-creation scenario in finite dimensions.  
As was reported by \cite{takayasu,takayasu2,krapivsky}
then the observed scaling $1/q^2$ is robust,
even when we confine the elements to diffusive
motion in 1 dimension, provided that creation of +/- pairs 
occur close to each other. For a visualization of the dynamic 
behaviour we in Fig.\ \ref{1d} show the evolving system in 1-d.
\\\\

\begin{figure}
\centerline{\epsfig{file=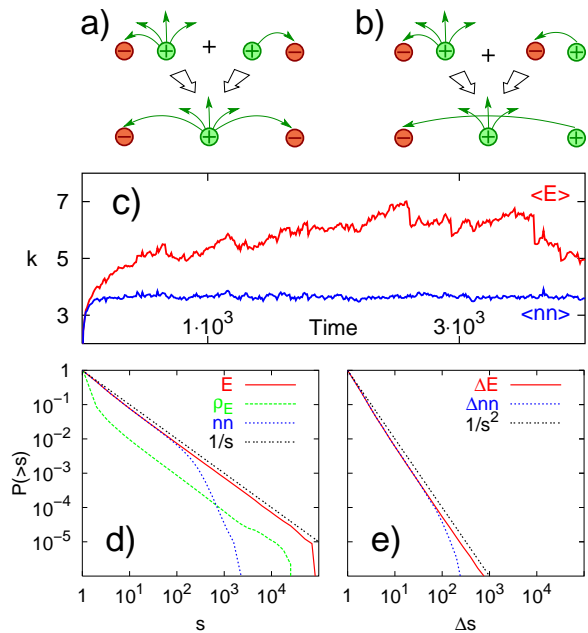, height=0.35\textheight,angle=0}}
\caption{The network realization of the model.
{\bf a)} and {\bf b)} illustrate possible merging moves. Positive vertices 
(donors) are vertices with outgoing edges and negative (acceptors) with 
incoming edges.
{\bf c)} The dynamics of the average number of edges per node, 
$\langle E \rangle$,
(upper curve) and the average number of neighbors $\langle nn \rangle$ 
(lower curve), $N=10^4$.
{\bf d)} The cumulative probability distributions, $N=10^5$, for: 
number of edges incoming or outgoing from a node, $E$ (solid curve);
number of neighbors, $nn$ (dotted curve);  
edge density, $\rho_E$ defined as the number of parallel edges 
connecting two vertices (dashed curve).
The distributions for all quantities are scale-free 
$P(>\!\!s)\sim 1/s^{\gamma-1}$ with $\gamma=2$.
{\bf e)} The cumulative probability distributions for the changes in 
number of edges due to merging, $\Delta E$ and number of 
neighbors $\Delta nn$.  The distributions are power-law 
$P(>\!\!\Delta s)\sim {\Delta s}^{1-\tau}$ with exponent 
$\tau = 2\gamma -1= 3$ from Eq.\ 4.}
\label{dynnet}
\end{figure}

We now consider a network implementation where each element is a vertex
and its sign corresponds to the number of in- or out-edges.
Thus the above scenario is translated to a network model in which donor
($q>0$) and acceptor ($q<0$) vertices are connected by
directed edges, see Fig.\ \ref{dynnet}(a) and (b). 
Each vertex may have different number of edges,
but at any time a given vertex cannot be both donor and acceptor.
Further, in the direct generalization of the model, we allow
several parallel edges between any pair of vertices. 
At each time-step two vertices $i$ and $j$ are chosen randomly. 
The update is then:
\begin{itemize}
\item 
Merge the two random vertices $i$ and $j$. There are now two
possibilities:\\ {\bf a)} 
If they have the same sign all the edges from $i$ and $j$
are assigned to the merged vertex.
Thereby the merged vertex has the same neighbors 
as $i$ and $j$ had together prior the merging, see Fig.\ \ref{dynnet}(a).\\
{\bf b)} If $i$ and $j$ have different signs, 
the resulting vertex is assigned the sign of
the sum $q_i+q_j$. Thereby a number $max\{ |q_i|, |q_j| \} -
|q_i+q_j|$ of edges are annihilated in such a way that 
only the two merging vertices change their number of edges.
This is done by reconnecting 
donor vertices of incoming edges to
acceptor vertices of outgoing edges, see Fig.\ \ref{dynnet}(b).

\item 
One new vertex is created of random sign, with one
edge being connected to a randomly chosen vertex.

\end{itemize}

On the vertex level this network model can be mapped to the above model
for merging and creation, and thus predict similar distributions
of vertex sizes, as seen by comparing solid line in Fig.\ \ref{dynnet}(d)
with Fig.\ \ref{pownet}(b) and distributions of annihilations in Fig.\ \ref{dynnet}(e) and Fig.\ \ref{pownet}(e).
However, the network formulation provides
additional insight into the excitation process that drives the
whole distribution. That is, starting with a number of empty
vertices $q_i=0$, the creation process generates vertex antivertex
pairs on small scale which subsequently may grow and shrink due to
merging and creation as illustrated in Fig.\ \ref{dynnet}(c).
One can see, that when the 
system has reached the stationary state, the average number of neighbors 
$\langle nn \rangle$ is nearly constant with 
small fluctuations while the fluctuations in 
the average number of edges, $\langle E \rangle $,
are much larger.  Further one notices that the
evolution of $\langle E \rangle $ is asymmetric, in 
the sense that increases are gradual, while decreases are
intermittent with occasional large drops in $\langle E \rangle$.
These drops primarily correspond to the merging of vertices
of different signs, where a large number of 
edges may be annihilated.
This process is quantified in Fig.\ \ref{dynnet}(e).

The network model opens
for a new range of power-laws \cite{hughes} 
associated to the connection pattern
and dynamics of reconnections between the vertices.
In this connection it is interesting that the 
number of edges per vertex, $E$, is distributed with scaling 
$P(E)\propto 1/E^2$. This was also obtained for the 
``number of loops at foot-point" in \cite{hughes}. 
In addition, the distribution of reconnection events 
$P_{\Delta}(\Delta E) \propto 1/\Delta E^3$ is distributed as 
the ``flare energies" in the model of Ref. \cite{hughes}.
In our model the event size is simply the change in the number
of edges ($\Delta E$) when two vertices merge which gives the exponent -3 as
shown in Eq.\ \ref{eq3}. 
In the model of Hughes et al. \cite{hughes} the event size is a more
complex quantity related to cascades of crossings of
field lines, and the energy release is associated to the 
number of lines that thereby decrease their length.
The non trivial fact that we obtain the same exponent
suggests that the two models are in the same 
universality class,
which means that our minimalistic model
captures the main features of a presumably 
much larger class of more detailed 
and realistic models.

Also we would like to mention that the distribution 
of the number of parallel edges for connected pairs of 
nodes is also scale invariant
$P(> \!\!\rho_E) \propto \rho_E^{1-\gamma}$ see Fig.\ \ref{dynnet}(d).
This illustrates robustness of the mechanism: 
The dynamics of merging vertices appears very different when it is
viewed from the ``dual" space of tubes of edges between vertices, $\rho_E$,
nevertheless the same exponent $\gamma=2$ is obtained.
\\\\

In conclusion we have discussed a new mechanism for obtaining 
scale-free networks of connected donor and acceptor vertices.
The model predicts power-laws of node degrees with a $1/s^2$ 
distribution, and of reconnection events with a $1/\Delta s^3$
distribution. 
The scenario thus provides a generic framework to
generate networks with large-scale features from small-scale excitations
under steady state conditions,
and may thus complement 
preferential growth which provides scaling
only under persistently growing conditions \cite{gronlund}.
Viewed as SOC, the merging-creation 
scenario provides ``scaling for free" in the sense that it 
is robust to multiple simultaneous updates. 
The key process of both constructive (equal sign) merging and destructive 
(opposite sign) merging \cite{ala} 
should be an important ingredient in a number of dynamic systems,
and in particular appears to be appealing minimalistic model
with possible connection to reconnection and creation of 
solar flares.


\begin{thebibliography}{10}
\bibitem{ala}
In fact even
$q_i\rightarrow q_i=|q_i\pm q_j|$ and $q_j\rightarrow q_j=1$
generates a distribution $p(q)\propto 1/q^2$ apart
from one continuously growing giant component that
takes care of the non-conservation in this particular update.
\end{thebibliography}
\end{document}